\documentclass[smallextended]{svjour3}
\smartqed  

\usepackage{amssymb}
\usepackage{amsmath}
\usepackage{graphicx}
\usepackage{color}

\usepackage{bbm}

\usepackage[utf8]{inputenc}

\usepackage{listings}

\newcommand{\mket}[1]{| #1 \rangle}
\newcommand{\mbra}[1]{\langle #1 |}

\newcommand{\mtr}[1]{\mathrm{Tr}\left( #1 \right)}

\newcommand{\nC}{\mathcal{C}}

\newcommand{\opd}[1]{\stackrel{#1}{\oplus}}

\newcommand{\imag}{\mathbbm{i}}

\begin{document}

\title{The Entanglement Level and the Detection of Quantum Data Transfer Correctness in Short Qutrit Spin Chains}


\author{Marek Sawerwain \and Joanna Wi\'sniewska
}



\institute{M.Sawerwain \at Institute of Control \& Computation Engineering \\
University of Zielona G\'ora, Licealna 9, Zielona G\'ora 65-417, Poland \\
\email{M.Sawerwain@issi.uz.zgora.pl}
\and
J.Wi\'sniewska \at Institute of Information Systems, Faculty of Cybernetics, \\
Military University of Technology, Kaliskiego 2, 00-908 Warsaw, Poland \\
\email{jwisniewska@wat.edu.pl}
} 

\date{Received: date / Accepted: date}

\maketitle

\begin{abstract}
The quantum entanglement is an important feature of many protocols in the field of quantum computing. In this paper we evaluate a level of entanglement in short qutrit chains. This evaluation is carried out with use of the CCNR criterion and the concurrence measure. We also present some explicit formulae describing the values of CCNR criterion and concurrence for exemplary short spin chains. Utilizing the obtained results, we indicate that analyzing the level of entanglement allows to detect the noise or deviation in the transfer process, in comparison to the perfect transfer where only operation realizing transfer is present.
\end{abstract}

\section{Introduction} \label{lbl:sec:introduction} 
A phenomenon of quantum entanglement \cite{Horodecki2009} is a fundamental feature of quantum systems. A presence of quantum entanglement in an examined system indicates system's quantum nature. The entanglement is also present in spin chains which became one of the essential elements of quantum physics and quantum computing \cite{Hirvensalo2001}, \cite{Nielsen2001}, \cite{Klamka2004}.

Using spin chains to transfer an information was shown for the first time in \cite{Bose2003}. An issue of the perfect transfer was widely discussed in \cite{Bose2007}, \cite{Kay2010}, \cite{Vinet2012}. An important problem of qubit and qudit states' transmission was raised in \cite{Amosov}, \cite{Bayat2014}, \cite{Jafarizadeh2008}. The next issues are: creating, detecting and analyzing the level of entanglement in two adjacent chain's nodes \cite{Almutairi2011}, \cite{Barasinski2014}. The questions connected with the presence of entanglement in spin chains are still an active research field. In \cite{Thomale2015}, \cite{Sahling2015}, \cite{Hao2007}, \cite{Zippilli2015} the analysis of entanglement in spin chains, also for qutrit chains, was presented.

The main function of spin chains is a transfer of an information. A level of entanglement in a spin chain changes during the transmission of quantum state/information. In spite of difficulties associated with formulating the entanglement levels, it is possible to evaluate the entanglement e.g. in spin chains. In this paper the level of entanglement is estimated for short qutrit spin chains. The presented results may be used to verify if the process of transfer is correct, because it seems that the level of entanglement can be treated as an invariant for the transfer protocol in a spin chain. 

The paper contains the following information: in Sect. (\ref{lbl:form:of:hamiltonian:for:qudits:MS:JW:CN:2015}) a form of Hamiltonian for performing a XY-like transfer protocol for qutrit and qudit chains is presented (including a short note concerning properties of the perfect transfer). There is also an algorithm describing the realization of transfer protocol with a $\gamma$ condition playing a role of the invariant expressing a level of entanglement during the transfer. In Sect. (\ref{lbl:entanglement:detection:and:measure:MS:JW:CN:2015}) a chosen criterion for entanglement detection and the concurrence measure are presented. 

Sect. (\ref{lbl:tracing:entanglement:MS:JW:after:CN2015:paper}) contains the results of the experiments for detecting entanglement and calculating the values of concurrence measure.

In Sect. (\ref{lbl:sec:transfer:paper:presence:MS:JW:after:CN2015:paper}) we show the influence of noise on the data transfer process. There is also an average value of Fidelity computed for a channel where distortions are presented by a phase-damping method.

A summary and conclusions are presented in Sect.~(\ref{lbl:sec:conslusions:MS:JW:after:CN2015:paper}).

\section{Hamiltonian for a Transfer Protocol in a Qudit Spin Chain}
\label{lbl:form:of:hamiltonian:for:qudits:MS:JW:CN:2015}

In this section we define a Hamiltonian $H^{{XY}_{d}}$ which will be used to realize the perfect transfer of quantum information in qutrit and qudit chains \cite{Jafarizadeh2008}, \cite{PazSilva2009}, \cite{Bayat2014} for entanglement creation between chosen points of a chain. The form of Hamiltonian, given below, is naturally suitable for transfers discussed in this work -- i.e. for transmission of information in qutrit spin chains.

The $SU(d)$ generator is utilized to create the XY-like Hamiltonian for qudits. The construction of suitable $SU(d)$ generator is given in Appendix A. Assuming that each qudit has the same freedom level $d \geq 2$ (the qudit is defined in a similar way to qubit, however a computational base for qudits is expressed with $d$ orthonormal vectors -- in case of qubits, the base contains two orthonormal vectors):

\begin{equation}
	H^{{XY}_{d}} = \sum_{(i,i+1) \in \mathcal{L}(G)} \frac{J_i}{2} \left( \Theta^{k,j}_{(i)} \Theta^{k,j}_{(i+1)} + \beta^{k,j}_{(i)} \beta^{k,j}_{(i+1)} \right), 
	\label{lbl:eqn:qudit:pst:hamiltonian:MSawe:JWisn:after:CN2015}
\end{equation}
where $J_i$ is defined as follows: $J_i = \frac{\sqrt{i(N-i)}}{2}$ for $1 \leq k < j < d$ and $\Theta^{k,j}_{(i)}$, $\beta^{k,j}_{(i)}$ are $SU(d)$ group operators defined by (\ref{lbl:eqn:theta:beta:operators:MSawe:JWisn:CN2014}) applied to the $(i)$-th and $(i+1)$-th qudit. The Hamiltonian (\ref{lbl:eqn:qudit:pst:hamiltonian:MSawe:JWisn:after:CN2015}) will be also called the transfer Hamiltonian.

The state transfers, studied in \cite{Bose2003}, \cite{Bose2007}, \cite{Vinet2012} use Hamiltonian $H$ which have the following property 
\begin{equation}
	[{H}, \sum_{i=1}^{N} Z_{(i)}] = 0,
\label{lbl:eqn:qubit:XY:Ham:comutator}	
\end{equation}
where $Z$ represents the sign gate for qubits. This means that spins are preserved and dynamics generated by $H$ is divided into series of subspaces denoted by the number of qubit in state $\mket{1}$ -- see in \cite{Kay2010}.
In the case discussed here, it is not hard to show that
\begin{equation}
	[H^{{XY}_{d}}, \sum_{i=1}^{N} \eta^{r,r}_{(i)}] = 0,
\label{lbl:eqn:qudit:Ham:comutator}
\end{equation}
for $1 \leq r \leq (d-1)$, so the equation (\ref{lbl:eqn:qudit:Ham:comutator}) generalizes the situation mentioned in the equation (\ref{lbl:eqn:qubit:XY:Ham:comutator}) -- preserving spins and separating dynamics into subspaces.

It is necessary to add that an appropriate unitary operator for transfer operation is determined by the equation
\begin{equation}
	U_t = e^{- \imag t H^{{XY}_{d}}},
\end{equation}
where $t$  represents evolution time and $\imag$ represents imaginary unity. The symbol $U$ represents the unitary operation which performs the transfer protocol. 

A very important issue raising in the context of information's transfer in spin chains is a problem of the perfect transfer. If the symbol $1$ denotes the initial node of the chain and $N$ stands for the final node then a transfer is the perfect transfer when for the time $t$ we have: 
\begin{equation}
|\mbra{N} e^{-i  H^{{XY}_{d}} t} \mket{1} | = 1 .
\end{equation}
In such a situation there were no distortions and the final and initial states are the same in the meaning of Fidelity measure.

Introducing a definition of the transfer protocol based on a Hamiltonian and a unitary operator allows to describe the transfer (Fig.~\ref{lbl:fig::MS:JW:CN:2015}) as an algorithm (or structural quantum program \cite{MSawerwainRGielerak2008}, \cite{Klamka2010}). A very important issue is the use of $\gamma$ condition as an invariant for the protocol. The invariant $\gamma$ is based on a function calculating the level of entanglement.

Let $E(\mket{\Psi(t)})$ be the function to determine a level of entanglement for a chain under perfect state transfer. The CCNR criterion, negativity and concurrence seem to be good candidates to play this role.

\begin{figure}
\begin{lstlisting}[mathescape]
$t$ ; time variable 
$N > 0$ ; number of step
$H^{{XY}_{d}}$ ; Hamiltonian for path $L$ with $l$ vertices)
$U_t = e^{-i (t/N) H^{{XY}_{d}}}$ ; unitary operator
$\mket{\psi(t_0)}$ ; initial state of chain

i:=0;

$\{ \gamma : E(\mket{\psi(t_i)}) \cong E(\mket{\Psi(t)}) \wedge i = 0\}$
while i < N do
begin 
    $\mket{\psi(t_{i+1})} := U_t \mket{\psi(t_i)}$
    i := i + 1
    $\{ \gamma : E(\mket{\psi(t_i)}) \cong E(\mket{\Psi(t)})\} \wedge i < N$
end;
$\{ \gamma : E(\mket{\psi(t_i)}) \cong E(\mket{\Psi(t)}) \wedge i \geq N \}$
\end{lstlisting} 
\caption{The data transfer process in a spin chain showed as an algorithm expressing the use of a unitary operator on a quantum register. The condition $\gamma$ plays the role of an invariant and describes the level of entanglement in the register $\mket{\psi(t_{i})}$ where $t_i$ represents the time in a step number $i$}
\label{lbl:fig::MS:JW:CN:2015}
\end{figure}

\begin{remark}
Calculating the level of entanglement in multiple qubit, qutrit and, particularly, qudit systems is a problem which is still unsolved for mixed quantum states (however it is possible to give lower bound for concurrence for mixed states \cite{Chen2005}). Due to this fact, in the notation of invariant $\gamma$ for perfect state protocol (algorithm) -- $\gamma : E(\mket{\psi(t_i)}) \cong E(\mket{\Psi(t)})$ -- sign $\cong$ expresses that the level of entanglement is comparable with used measure (e.g. CCNR criterion or concurrence). 
\end{remark}

In this paper the transfer protocol is used to realize the transmission of information through the path which length equals $N$. In general the process of transfer involves an unknown qudit state (where $d$ is a freedom level):
\begin{equation}
	\mket{\psi} = \alpha_0 \mket{0} + \alpha_1 \mket{1} + \ldots + \alpha_{d-1} \mket{d-1} \;\;\;\ \mathrm{where} \; \alpha_i \in \mathbb{C} \; \mathrm{and} \;  i < d .
\end{equation}
However, in our examples described in further part of this work, we transfer only the qutrit state, i.e. $d=3$.

\begin{figure}
	\begin{tabular}{c}
		\includegraphics[height=0.65cm]{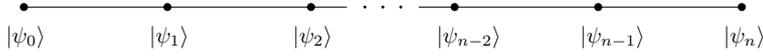} 
	\end{tabular}
	\centering
	\caption{The picture of quantum information transfer's realization  in a spin chain for a single unknown qudit state $\mket{\psi}$ in the path of length $l$. The state $\mket{\psi_0}$ represents the initial state and state $\mket{\psi_n}$ denotes the final state obtained after execution of transfer protocol}
	\label{lbl:fig:transfer:information:in:qudits:chain:path:MS:JW:CN:2015}
\end{figure}

Fig. \ref{lbl:fig:transfer:information:in:qudits:chain:path:MS:JW:CN:2015} shows the scheme of transfer protocol's realization. The whole process is divided into discrete steps which are realized by operator $U_t$. 

\begin{figure}

\begin{tabular}{c}
	\includegraphics[height=4.0cm]{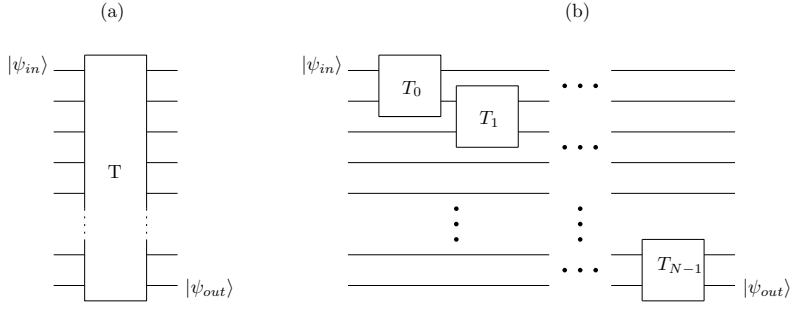}
\end{tabular}

\centering

\caption{The examples of quantum circuits realizing the transfer protocol for $N$ nodes. In the case (a) we assume that the whole process will be realized by one unitary operator $T$, so this operation needs an access to all chain's nodes. The case (b) shows the quantum circuit where the transfer process is realized by local operators $T_i$ correlated with two adjoining chain's nodes}

\label{lbl:fig:transfer:information:in:qudits:circuit:MS:JW:CN:2015}
\end{figure}

\section{The CCNR Criterion and Concurrence} \label{lbl:entanglement:detection:and:measure:MS:JW:CN:2015}

In this part a basic information about CCNR criterion and concurrence will be recalled. The mentioned methods may be used as measures to estimate a level of entanglement in a bipartite system. The CCNR criterion uses two auxiliary operations called vectorization and realignment.

The vectorization of a given matrix $A = [a_{ij}]_{m \times n}$ is represented by the operation $vec: A_{m \times n} \rightarrow \mathbb{C}^{mn}$, and
\begin{equation}
	vec(A) = [ a_{11}, \ldots, a_{m1}, a_{12}, \ldots, a_{m2}, \ldots,  a_{1n}, \ldots, a_{mn} ]
\end{equation}

The realignment is based on the vectorization of matrix and applies to the scalar product of two matrices: 
\begin{equation}
	R( A \otimes B ) = {vec(A)}^{T} \cdot vec(B) .
\end{equation}
The realignment of matrix's elements -- the initial matrix dimensions are $2 \times 2$ -- in a space $H = H_A \otimes H_B$ may be illustrated by the example:
\begin{equation}
	\rho=\left(%
	\begin{array}{cc|cc}
		a_{11} & a_{12} & a_{13} & a_{14} \\
		a_{21} & a_{22} & a_{23} & a_{24} \\ \hline
		a_{31} & a_{32} & a_{33} & a_{34} \\
		a_{41} & a_{42} & a_{43} & a_{44} \\
	\end{array}%
	\right), \;\;\;
	R(\rho)=\left(%
	\begin{array}{cc|cc}
		a_{11} & a_{21} & a_{12} & a_{22} \\
		a_{31} & a_{41} & a_{32} & a_{42} \\ \hline
		a_{13} & a_{23} & a_{14} & a_{24} \\
		a_{33} & a_{43} & a_{34} & a_{44} \\
	\end{array}%
	\right)
\end{equation} 

The operation of realignment is realized similarly for larger matrices, e.g. for matrices $3 \times 3$: 
\begin{equation}
	R(\rho)=\left(%
	\begin{array}{ccc|ccc|ccc}
		a_{11} & a_{21} & a_{31} & a_{12} & a_{22} & a_{32} & a_{13} & a_{23} & a_{33} \\
		a_{41} & a_{51} & a_{61} & a_{42} & a_{52} & a_{62} & a_{43} & a_{53} & a_{63} \\ 
		a_{71} & a_{81} & a_{91} & a_{72} & a_{82} & a_{92} & a_{73} & a_{83} & a_{93} \\ \hline
		
		a_{14} & a_{24} & a_{34} & a_{15} & a_{25} & a_{35} & a_{16} & a_{26} & a_{36} \\
		a_{44} & a_{54} & a_{64} & a_{45} & a_{55} & a_{65} & a_{46} & a_{56} & a_{66} \\ 
		a_{74} & a_{84} & a_{94} & a_{75} & a_{85} & a_{95} & a_{76} & a_{86} & a_{96} \\ \hline
		
		a_{17} & a_{27} & a_{37} & a_{18} & a_{28} & a_{38} & a_{19} & a_{29} & a_{39} \\
		a_{47} & a_{57} & a_{67} & a_{48} & a_{58} & a_{68} & a_{49} & a_{59} & a_{69} \\ 
		a_{77} & a_{87} & a_{97} & a_{78} & a_{88} & a_{98} & a_{79} & a_{89} & a_{99} 
	\end{array}%
	\right)
\end{equation} 

The realignment allows to specify the criterion which is able to detect and track the entanglement's behaviour during the transfer process. This method is called CCNR criterion \cite{Rudolph2003a}. Generally, the CCNR criterion is defined by the fact that if the matrix $\rho_{AB}$ of a~bipartite $m \times n$ system is separable, then:
\begin{equation}
	|| \rho^R_{AB} || \leq 1 ,
	\label{lbl:thm:cross:norm}
\end{equation}
in the case $|| \rho^R_{AB} || > 1$ state $\rho_{AB}$ is entangled. Of course $\rho^R$ stands for the completion of realignment operation on the state $\rho$. 

\begin{remark}
Generally, the CCNR criterion is not sufficient to detect all the cases of entanglement in any quantum system. However, it is sufficient to correctly detect the entanglement in qudit, and of course qutrit, spin chains.
\end{remark}

It should be also pointed out that the value of CCNR criterion may be calculated using Singular Value Decomposition (SVD):
\begin{equation}
|| \rho^R_{AB} || = \sum_{i=1}^{q} \sigma_i (\rho_{AB}^R)
\label{lbl:thm:cross:norm:with:svd}
\end{equation}
where $\sigma_i$ represents a~singular value of $\rho_{AB}^R$ and $q=\min(m^2,n^2)$.

The correctness of CCNR criterion may be amplified by using the states of subsystems what was shown in \cite{Zhang2008}:  
\begin{equation}
|| (\rho_{AB} - \rho_A \otimes \rho_B)^{R} || \leq \sqrt{(1-\mtr{\rho^2_A})(1-\mtr{\rho^2_B})} ,
\label{lbl:thm:stronger:cross:norm:MS:JW:CN:2015}
\end{equation}

If the above condition is fulfilled, it means that the state $\rho_{AB}$ is separable (otherwise the state is entangled).

The concurrence measure is used to calculate a level of entanglement, as well. Concurrence \cite{Chen2005} for pure state $\mket{\psi_{AB}}$ in bipartite system $d \otimes d$ is defined as
\begin{equation}
	\nC( \mket{\psi_{AB}} ) = \sqrt{ 2 \left( 1 - \mtr{\rho^2_A} \right) } .
	\label{lbl:eqn:concurrence:MS:JW:after:CN:2015}
\end{equation}

\section{Tracing of Entanglement} \label{lbl:tracing:entanglement:MS:JW:after:CN2015:paper}

\begin{figure}[!ht]
	\begin{tabular}{c}
		\includegraphics[width=12cm]{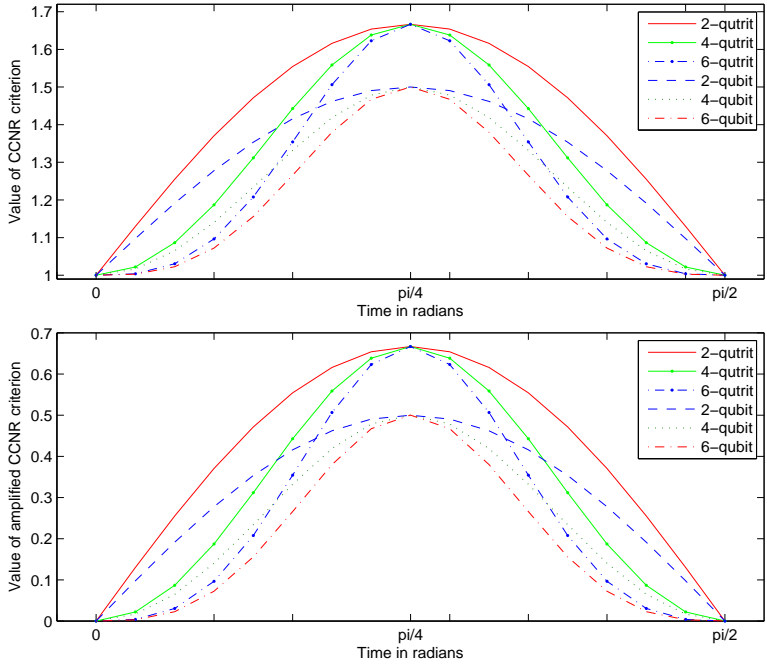}
	\end{tabular}
	\centering
	\caption{The entanglement detection with use of CCNR criterion and amplified CCNR criterion for short qubit and qutrit chains. The presented values were calculated numerically and the transferred state was $\mket{+}$ state}
	\label{lbl:fig:entanglement:level:ccnr:qudits:chain:path::MS:JW:CN:2015}
\end{figure}

In this section we present the results of calculating a level of entanglement with use of CCNR criterion and concurrence. Fig.~\ref{lbl:fig:entanglement:level:ccnr:qudits:chain:path::MS:JW:CN:2015} shows the values of CCNR criteria calculated according to the equations (\ref{lbl:thm:cross:norm:with:svd}) and (\ref{lbl:thm:stronger:cross:norm:MS:JW:CN:2015}). The process of transfer was realized for the state $\mket{+}$, which for qubits is $\mket{+}_2$ state and for qutrits is expressed as $\mket{+}_3$ state:
\begin{equation}
	\mket{+}_2 = \frac{1}{\sqrt{2}}(\mket{0} + \mket{1}) \;\;\; or \;\;\;\  \mket{+}_3 = \frac{1}{\sqrt{3}}(\mket{0} + \mket{1} + \mket{2}).
\end{equation}

The value of CCNR criterion was calculated according to the equation (\ref{lbl:thm:cross:norm:with:svd}). In case of amplified CCNR criterion, the equation (\ref{lbl:thm:stronger:cross:norm:MS:JW:CN:2015}) was transformed to:
\begin{equation}
	|| (\rho_{AB} - \rho_A \otimes \rho_B)^{R} || - \sqrt{(1-\mtr{\rho^2_A})(1-\mtr{\rho^2_B})} > 0.
\end{equation}
If the above inequality is true, then the analysed state is entangled.

Naturally, during the transfer process, the level of entanglement initially increases and then decreases. All analysed short spin chains were divided into two subsystems: $A$ and $B$. 
\begin{remark}
It should be pointed out that the subsystems $A$ and $B$ are treated as systems with a greater number of dimensions. However, it is still possible to detect the entanglement with use of CCNR criterion and to evaluate the level of entanglement with concurrence.
\end{remark}

\begin{figure}
	\begin{tabular}{c}
		\includegraphics[width=10cm]{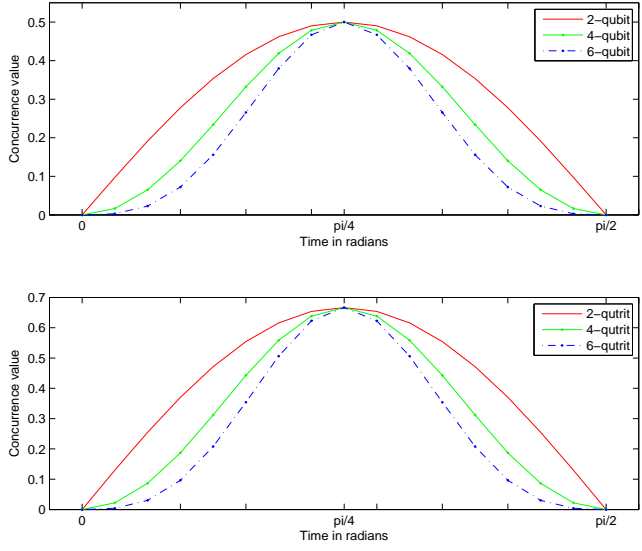}
	\end{tabular}
	\centering
	\caption{The values of concurrence measure for short qubit and qutrit chains. The presented results were calculated numerically and the transferred state was $\mket{+}$ state}
	\label{lbl:fig:entanglement:level:concurrence:qudits:chain:path::MS:JW:CN:2015}
\end{figure}

Fig.~\ref{lbl:fig:entanglement:level:concurrence:qudits:chain:path::MS:JW:CN:2015} shows the values of concurrence measure for some exemplary chains. It can be observed that the entanglement's dynamics is similar to the results from Fig.~\ref{lbl:fig:entanglement:level:ccnr:qudits:chain:path::MS:JW:CN:2015}.  

Using the equation (\ref{lbl:eqn:concurrence:MS:JW:after:CN:2015}) allows to present the explicit formulae for calculating the level of entanglement in spin chains. If an unknown qubit state: 
\begin{equation}
	\mket{ \psi } = \alpha \mket{0} + \beta \mket{1},
\end{equation}
is transferred through the short path ($l=2$), then the level of entanglement in this 2-qubit chain can be expressed as:
\begin{equation}
	\nC^{l=2}_{d=2}(a) = \frac{1}{4} \left( 4 \alpha ^4 +3 \beta ^4 +  8 \alpha ^2 \beta ^2 \cos (2 a)+\beta ^4 \cos (4 a) \right)
	\label{lbl:eqn:direct:formula:2qubits}
\end{equation}
where $a$ represents the duration of the transfer process.

For a 2-qutrit chain the transfer of an unknown state:
\begin{equation}
	\mket{ \psi } = \alpha \mket{0} + \beta \mket{1} + \gamma \mket{2},
\end{equation}
can be precisely described by the concurrence:
\begin{gather}
	\nC^{l=2}_{d=3}(a) = \frac{1}{4} \left(4 \alpha^4 + 3(\beta^2+\gamma^2)^2 + 8 \alpha ^2 \left(\beta ^2+\gamma ^2 \right) \cos(2 a) + \left(\beta^2+\gamma ^2\right)^2 \cos(4a) \right)
	\label{lbl:eqn:direct:formula:2qutrits}
\end{gather}

\begin{figure}[!ht]
	\begin{tabular}{c}
		\includegraphics[width=12cm]{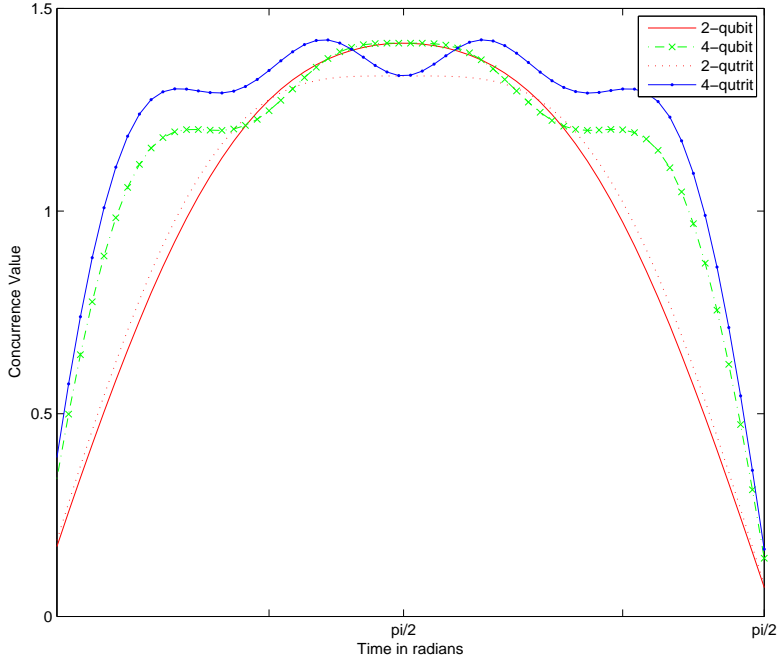}
	\end{tabular}
	\centering
	\caption{The values of concurrence measure, for short spin chains, obtained with use of the formulae (\ref{lbl:eqn:direct:formula:2qubits}) and (\ref{lbl:eqn:direct:formula:2qutrits})}
	\label{lbl:fig:concurrence:level:direct:formula:for:chain:path::MS:JW:CN:2015}
\end{figure}

The dynamics of changes for concurrence values depends on the values $\cos(2a)$ and $\cos(4a)$ -- both in qubit and qutrit chains. The constants preceding $\cos(2a)$ and $\cos(4a)$ correspond to the probability amplitudes. A sum of probability amplitudes' values equals one what is easy to proof when for some $a$ values: $\cos(2a)=\cos(4a)=1$. Furthermore the values of amplitudes:
\begin{equation}
	(\alpha^2 + \beta^2)^2 = 1, \;\;\; (\alpha^2 + \beta^2 + \gamma^2 )^2 = 1
\end{equation}
both for qubit and qutrit chains of any length, because of a trace calculation in~ (\ref{lbl:eqn:concurrence:MS:JW:after:CN:2015}). An exemplary value of concurrence measure for 4-qutrit chain is:
\begin{gather}
	\nC^{l=4}_{d=3}(a) = \frac{1}{64} \left( c_0 + 
	(c_1 +  c_2) \cos 2a  + 
	(c_3  + c_4 + c_5) \cos 4a + \right. \\ \notag
	\left. (c_6  +  c_7) \cos 6a + 
	(c_8 +  c_9  +  c_{10}) \cos 8a + 
	(c_{11} +  c_{12}   + c_{13}) \cos 12a \right)
\end{gather}
where $\sum_{i} c_i = 1$ and the $c_i$ coefficients are composed of the probability amplitudes' values of transferred state. 

Fig.~\ref{lbl:fig:concurrence:level:direct:formula:for:chain:path::MS:JW:CN:2015} presents the values of concurrence measure for explicit formulae. The changes in the level of entanglement, illustrated as local minima, were not detectable with numerical calculations (see Fig. \ref{lbl:fig:entanglement:level:ccnr:qudits:chain:path::MS:JW:CN:2015}). These minima correspond to the different levels of entanglement between separate chain's parts, e.g. between adjacent qubits/qutrits.

\begin{remark}
It should be noted that usually there is no need to analyze the global level of entanglement for the whole chain. The character of a transfer in a one-dimensional spin chain causes that the entire transferred information will be placed in the last node in the last step of the process. It means that the levels of entanglement can be calculated only in the first and the last node to check if the transfer process is correct. The needed quantum state, for the spin chain with the number of nodes greater than two, we obtain by use of the partial trace operation. If $\rho$ represents the state of the whole chain with $N$ nodes then the state in the first (denoted as $1$) and the last (described as $N$) node we calculate as:
\begin{equation}
\rho_{1N} = \mathrm{Tr}_{(2,3,...,i-k, \ldots, N-2, N-1)}(\rho)
\end{equation}
where $\mathrm{Tr}_{( \cdot )}(\rho)$ is the partial trace operation.
\end{remark}

\section{Transfer with Noise Presence} \label{lbl:sec:transfer:paper:presence:MS:JW:after:CN2015:paper}

Naturally, the presence of noise in a transfer process causes a distortion of quantum state what can be verified, i.a., with use of Fidelity measure. The presence of noise may be analyzed in the context of Grover's algorithm \cite{Gawron2012} and usually the distortions severely hamper the realization of a quantum algorithm or a quantum protocol.

However, calculating the value of Fidelity needs the information about the quantum state before and after the transfer. We know that the initial state is separable and the phenomenon of entanglement should only be present during the transfer process (if the transfer is perfect, the final state is also separable). This issue should make us reflect that calculating the level of entanglement for the final state, with use one of known measures, may tell us if the transfer was perfect. 

The concurrence is relatively easy to compute so we will use this measure (\ref{lbl:eqn:concurrence:MS:JW:after:CN:2015}) to detect the presence of noise. 

The distortions will be presented as a phase-damping in a qutrit spin chains, expressed as a quantum channel in a Kraus representation. Phase-damping will be applied to the whole qutrit chain. However, for qutrits, and for qudits in general, the phase-damping operation does not have a unique representation. The model discussed in the works \cite{Fukuda}, \cite{LiuEtAll}, \cite{Amosov} is one of the examples of this operation. In our work we use the following operators for phase-damping:

\begin{equation}
	\mathcal{E}(\rho) = \sum_{i=0}^{d-1} E_i \rho E^{\dagger}_i, \;\;\; E_i = \sqrt{ {{d-1} \choose {i}} {\left( {1-p} \over 2 \right)}^i {\left( {1+p} \over 2 \right)}^{d-1-i} } Z^i ,
	\label{MSawe:JWisn:CN2013:lbl:eq:phase:damping:qudit}
\end{equation}
where the damping strength is as follows $0 \leq p \leq 1$ and ${{d-1} \choose {i}}$ represents the binomial theorem. 

\begin{remark}
	It should be noted that expression (\ref{MSawe:JWisn:CN2013:lbl:eq:phase:damping:qudit}) can be regarded as a~special case of Weyl's channel \cite{Fukuda}:
	\begin{equation}
		\mathcal{E}(\rho) = \sum\limits_{m,n = 0}^{d-1} \pi_{m,n} (Z^{n}X^{m}) \rho {(X^{m} Z^{n})}^{\dagger} ,
	\end{equation}
	where elements of the matrix $\pi$ satisfy the following conditions: $0 \leq \pi_{m,n} \leq 1$ and $\sum_{m,n=0}^{d-1} \pi_{m,n} = 1$. The operators $Z$ and $X$ are generalised Pauli matrices for the sign changing and negation operations on qudits.
\end{remark}

The definitions of qudit gates $X$ and $Z$ are:
\begin{equation}
X \mket{j} = \mket{ j \opd{d} 1 }, \;\;\; Z \mket{j} = \omega^{j}\mket{j},
\label{lbl:eq:pauli:qutrit}
\end{equation}
where $\omega$ stands for $k$-th root of unity:
\begin{equation}
\omega^d_k = \cos \left( \frac{2k\pi}{d} \right) + \imag \sin\left(\frac{2k\pi}{d}\right) = e^{\frac{2\pi \imag k}{d}}, \;\;\; k=0,1,2,\ldots,d-1,
\end{equation}
$d$ expresses the degree of root and $\imag$ denotes the imaginary unit. It is also assumed that
\begin{equation}
j \opd{d} a = (j+a) \; \mathrm{mod} \; d, \;\;\; j \in N.
\end{equation}

\begin{figure}
	\begin{tabular}{c}
		\includegraphics[height=7.0cm]{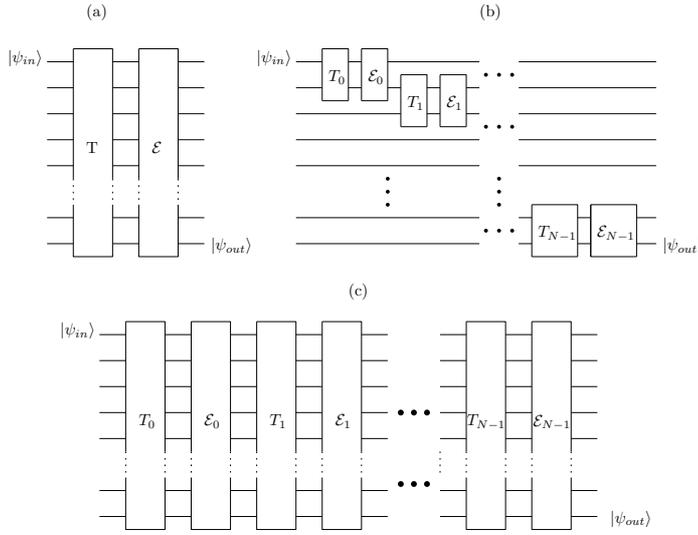}
	\end{tabular}
	\centering
	\caption{The diagrams of quantum circuits realizing the transfer of quantum state. In a case (a) the transfer is realized with use of one $T$ gate (this operation is performed in every node) and the distortion is presented as an $\mathcal{E}$ gate. A case (b) expresses the situation where a transfer and a distortion are realized locally (effecting only adjacent nodes). In a case (c) the operators $T$ and $\mathcal{E}$  are decomposed to a sequence of $T_i$ and $\mathcal{E}_i$ operations which are performed alternately}
	\label{lbl:fig:transfer:information:circuit:with:noise:MS:JW:after:CN:2015}
\end{figure}

The distortion operator will be utilized after the operation of data transfer. The transfer operator is a global operator which affects all the elements in a short spin chain. Three basic ways to introduce the noise to a spin chain are presented at Fig.~(\ref{lbl:fig:transfer:information:circuit:with:noise:MS:JW:after:CN:2015}).

\begin{figure}
	\begin{tabular}{c}
		\includegraphics[height=7.0cm]{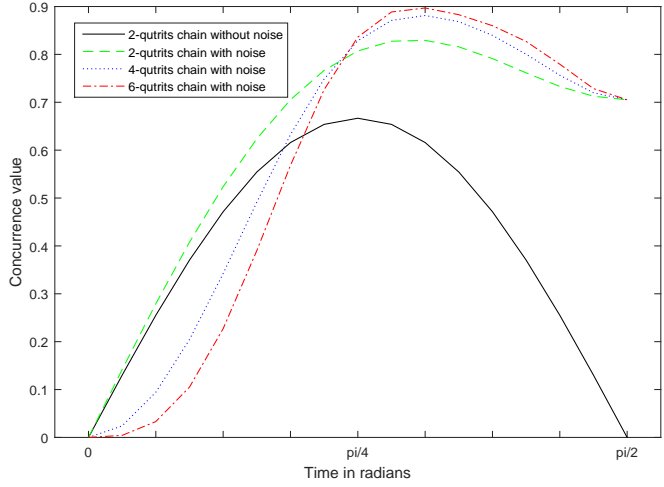}
	\end{tabular}
	\centering
	\caption{The value of concurrence measure during the data transfer in two and more qutrit chains. If the noise is present, it is generated as a phase-damping with parameter $p = 0.85$}
	\label{lbl:fig:concurrency:noise:transfer:information:MS:JW:after:CN:2015}
\end{figure}

Fig.~\ref{lbl:fig:concurrency:noise:transfer:information:MS:JW:after:CN:2015} depicts the values of concurrence measure for a few experiments with short spin chains. The diagram shows that the presence of noise significantly changes the level of entanglement during the transfer process (the level of entanglement should decrease in the final part of the process). The distortions prevent the drop of entanglement's level, so the presence of entanglement in the final state means that the transfer was not completed successfully. 

\begin{figure}
	\begin{tabular}{c}
		\includegraphics[height=4.0cm]{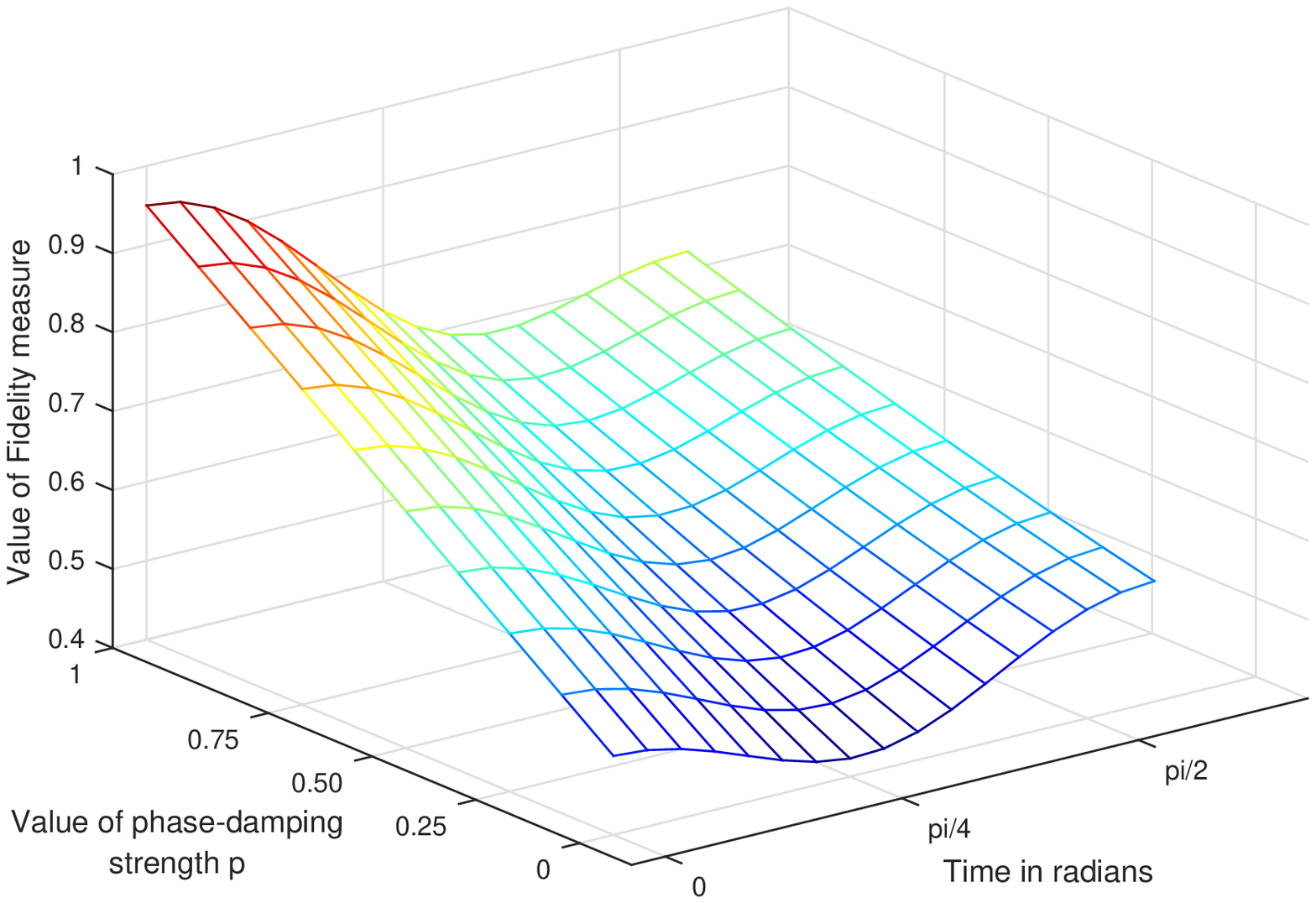}
		\includegraphics[height=4.0cm]{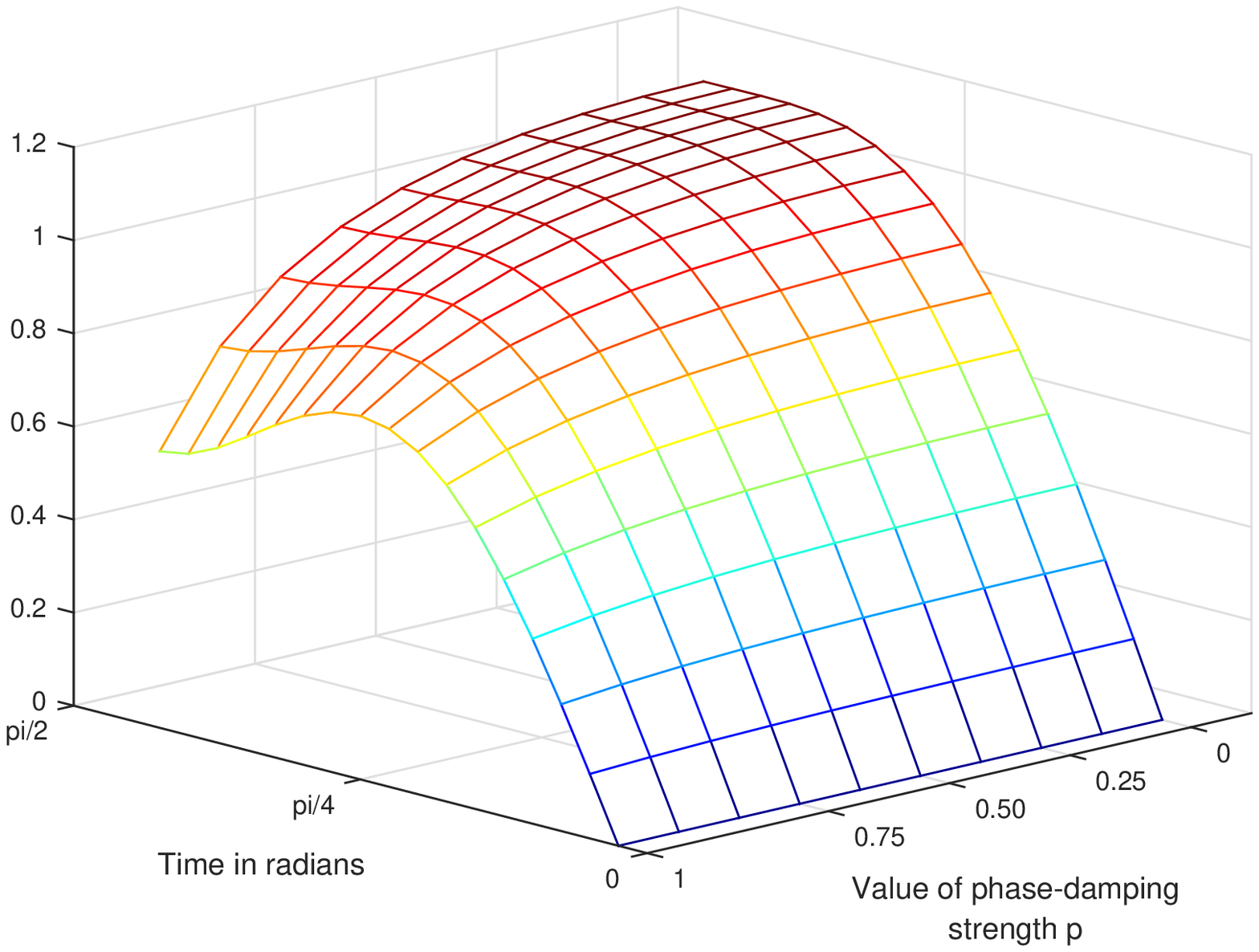}
	\end{tabular}
	\centering
	\caption{The values of Fidelity and Concurrence measures for the 2-qutrit chain where the transfer process was divided into sixteen steps and the distortions were caused by phase-damping (with different parameters)}
	\label{lbl:fig:fidelity:noise:transfer:information:MS:JW:after:CN:2015}
\end{figure}

We can confirm this result also by calculating the average value of Fidelity for the perfect transfer and transfer operators in e.g. 2-qutrit chains. The works \cite{Nielsen2002}, \cite{Pedersen2007}, \cite{Bowdrey2002} show that the average value of Fidelity may be computed as:
\begin{equation}
F_{avg}(\psi, U_0, \mathcal{E}) = \frac{1}{n(n+1)} \left( \mtr{\sum_k M^{\dagger}_{k}M_k} + \sum_k {| \mtr{M_k} |}^2 \right)
\end{equation}
where $M_k = {U_t}^{\dagger} E_k$ and $\mathcal{E}(\rho) = \sum_{k} E_k \rho E^{\dagger}_k$ represent the quantum channel -- in this case these are phase-damping operators. The unitary operator $U_t$ realizes the perfect transfer. The average value of Fidelity, calculated analytically, for a 2-qutrit spin chain:
\begin{equation}
F_{avg} = \frac{1}{15} \left(3p^2+{\left|{p}^2-1\right|}+4{p}+3\right).
\end{equation}
The average value of Fidelity equals $2/3$ what is confirmed by our numerical experiment (see Fig.~(\ref{lbl:fig:fidelity:noise:transfer:information:MS:JW:after:CN:2015})) -- the average value of Fidelity, calculated numerically, is $F_{avg} \approx 0.62702$.

At Fig.~(\ref{lbl:fig:fidelity:noise:transfer:information:MS:JW:after:CN:2015}) the level of entanglement is also presented. Its value confirms that the transfer process did not end correctly.

\section{Conclusions} \label{lbl:sec:conslusions:MS:JW:after:CN2015:paper}

Calculating the level of entanglement with use of CCNR criterion and, especially, with concurrence proved to be a good solution for tracking the information transfer in qutrit spin chains. The concurrence measure indicates unambiguously the details of transfer process and due to this fact, it can be used as an invariant in the algorithmic description of transfer protocol for an unknown qubit state and also for an unknown qutrit state (generalization for qudit states is also available).

Naturally, the presented analysis concerns only short spin chains and it should be extended for chains of any length. However, this extension is in fact only estimating the upper bound for concurrence value in a spin chain. The numerical experiments seem to imply that the length $l=2$ points out the upper bound for concurrence in a spin chain given by the Hamiltonian (\ref{lbl:eqn:qudit:pst:hamiltonian:MSawe:JWisn:after:CN2015}).

Introducing the noise in a form of phase-dumping causes a higher level of entanglement and a lower value of the Fidelity measure. Tracking the entanglement level in chains with and without noise allows to study the distortions. We suppose that analyzing the level of entanglement in chains with and without noise, for example to compare the values of Fidelity measure, would be an interesting research issue.

It is important to add that the concurrence measure -- apart from its analytical functions, e.g. to track the level of entanglement, as it was shown in the paper -- is at present also used in the experiments \cite{Almutairi2011} where the entanglement is produced in a two-photon system. In \cite{Wang2009} the concurrence was utilized for spin-1/2 chain which can be realized physically with use of Benzene molecules \cite{Bose2003}. Therefore, the functions described in the chapter may be applied as a tool to verify the concurrence in the physical experiments where e.g. qutrits were used.

\begin{acknowledgements}
We would like to thank for useful discussions with the~\textit{Q-INFO} group at the Institute of Control and Computation Engineering (ISSI) of the University of Zielona G\'ora, Poland. We would like also to thank to anonymous referees for useful comments on the preliminary version of this paper. The numerical results were done using the hardware and software available at the ''GPU $\mu$-Lab'' located at the Institute of Control and Computation Engineering of the University of Zielona G\'ora, Poland. 
\end{acknowledgements}

\section*{Appendix A -- Construction of Lie algebra's generator}

In a following definition of the XY-like Hamiltonian for qutrits' and, generally, qudits' chain, the Lie algebra's generator for a group $SU(d)$ was used, where $d \geq 2$ is to define a set of operators responsible for transfer dynamics. For clarity, the construction procedure of $SU(d)$ generators will be recalled -- in the first step a set of projectors is defined:

\begin{equation}
(P^{k,j})_{\upsilon, \mu} = \mket{k}\mbra{j} = \delta_{\upsilon, j} \delta_{\mu, k},  \;\;\; 1 \leq \upsilon, \mu \leq d .
\end{equation}
The first suite of $d(d-1)$ operators from the group $SU(d)$ is specified as 
\begin{equation}
\Theta^{k,j}  =  P^{k,j} + P^{j,k}, \;\;\; \beta^{k,j}  = -i (P^{k,j} - P^{j,k}),
\label{lbl:eqn:theta:beta:operators:MSawe:JWisn:CN2014}
\end{equation}
and $1 \leq k < j \leq d$.

The remaining $(d-1)$ generators are defined in the following way 
\begin{equation}
\eta^{r,r} = \sqrt{\frac{2}{r(r+1)}} \left[ \left( \sum^{r}_{j=1} P^{j,j} \right) - r P^{r+1,r+1} \right],
\end{equation}
and $1 \leq r \leq (d-1)$. Finally, the $d^2-1$ operators belonging to the $SU(d)$ group can be obtained.

\end{document}